\documentclass[twocolumn]{aastex62}
\usepackage{amsmath}     
\usepackage{wasysym}           
\usepackage{graphicx}
\usepackage{amssymb}
\usepackage{epstopdf}
\usepackage{mathrsfs}
\usepackage{anyfontsize}
\usepackage{natbib}
\usepackage{color}
\usepackage{lipsum}
\usepackage{diagbox}

\shorttitle{Massive Black Hole Growth Inside Quasi-stars}
\shortauthors{Coughlin \& Begelman}
\begin{document}
\title{Quasi-stars as a Means of Rapid Black Hole Growth in the Early Universe}
\author[0000-0003-3765-6401]{Eric R.~Coughlin}
\affiliation{Department of Physics, Syracuse University, Syracuse, NY 13210, USA}
\author[0000-0003-0936-8488]{Mitchell C.~Begelman}
\affiliation{JILA, University of Colorado and National Institute of Standards and Technology, 440 UCB, Boulder, CO 80309-0440, USA} 
\affiliation{Department of Astrophysical and Planetary Sciences, University of Colorado, 391 UCB, Boulder, CO 80309-0391, USA}

\email{ecoughli@syr.edu}

\begin{abstract}
JWST observations demonstrate that supermassive black holes (SMBHs) exist by redshifts $z \gtrsim 10$, providing further evidence for ``direct collapse'' black hole (BH) formation, whereby massive ($\sim 10^{3-5} M_{\odot}$) SMBH seeds are generated within a few Myr as a byproduct of the rapid inflow of gas into the centers of protogalaxies. Here we analyze the intermediate ``quasi-star'' phase that accompanies some direct collapse models, during which a natal BH accretes mass from and energetically sustains (through accretion) an overlying gaseous envelope. We argue that previous estimates of the maximum BH mass that can be reached during this stage, $\sim 1\%$ of the total quasi-star mass, are unphysical, and arise from underestimating the efficiency with which energy can be transported outward from regions close to the BH. We construct new quasi-star models that consist of an inner, ``saturated-convection’’ region (which conforms to a convection-dominated accretion flow near the BH) matched to an outer, adiabatic envelope. These solutions exist up to a BH mass of $\sim 60\%$ the total quasi-star mass, at which point the adiabatic envelope contains only 2\% of the mass (with the remaining $\sim 38\%$ in the saturated-convection region), and this upper limit is reached within a time of $20-40$ Myr. We conclude that quasi-stars remain a viable route for producing SMBHs at large redshifts, consistent with recent JWST observations.
\end{abstract}


\section{Introduction}
The means by which quasars, which host supermassive black holes (SMBHs) with mass in excess of $\sim 10^{9}M_{\odot}$, came to exist at $z \gtrsim 5-8$ \citep{fan03, mortlock11, wu15, banados18, lai24}, is still debated. If one maintains accretion at the Eddington rate, one can just grow a stellar-mass black hole (BH) to $10^{9}M_{\odot}$ by $\sim 1$ Gyr following the Big Bang. However, how this accretion rate could be sustained for millions to billions of years is unclear. Instead, it seems likely that these idealized conditions are simply not manifested, and some other process is at work (at least in tandem with relatively steady accretion) to produce SMBHs in the early Universe. 

This conclusion --- that SMBHs grow at a hyper-Eddington rate for a significant amount of time --- seems even more inescapable given the recent discoveries by JWST, which demonstrate not only that many massive galaxies have formed by $z \gtrsim 10$ (e.g., \citealt{naidu22, labbe23, castellano22, castellano23}), but also that there are likely many active galactic nuclei (AGN) at $z \gtrsim 7$ (e.g., \citealt{harikane23}). X-ray follow-up observations of gravitationally lensed galaxies have directly confirmed the existence of a massive ($M_{\bullet} \gtrsim 10^{7-8}M_{\odot}$, with $M_{\bullet}$ the mass of the SMBH) and actively accreting SMBH at $z \gtrsim 10$ \citep{bogdan24, natarajan24}, which is $\lesssim 500$ Myr after the Big Bang. The latter detection is particularly suggestive as to the nature of early BH growth, as the ratio of the mass of the SMBH (assuming the current accretion rate is Eddington; sub-Eddington accretion only increases the ratio) to the stellar mass of the galaxy in which it resides is $\sim 10\%$ --- significantly higher than the values of $0.1 - 1\%$ measured at $z \simeq 0$ (e.g., \citealt{reines15}). This relatively high ratio of BH to stellar mass is consistent with ``heavy seed'' formation channels, in which the first BHs to form had masses in the intermediate-mass black hole (IMBH) range ($10^{3} M_{\odot} \lesssim M \lesssim 10^{5} M_{\odot}$) and collapsed directly or nearly directly from relatively warm and primordial gas clouds (e.g., \citealt{haehnelt93, umemura93, loeb94, eisenstein95a, madau01, bromm03, begelman06, lodato06, volonteri08, hosokawa11, alexander14, shlosman16, natarajan17, pacucci17, wise19}). The ``direct collapse'' means of high-mass BH formation can be contrasted with models in which SMBHs formed hierarchically (and through subsequent accretion) from the remnants of Pop.~III stars (e.g., \citealt{haiman01, volonteri03, yoo04, volonteri05, sesana09, tanaka09, whalen12}; see also \citealt{natarajan12, natarajan14} for comparisons of the two channels).

Direct collapse models come in several forms (e.g., the early work by \citealt{begelman78} suggests a number of possibilities).  Assuming that the necessary conditions are present to avoid the fragmentation of gas into stars (generally temperatures $\gtrsim 10^{4}$ K and low metallicity; \citealt{oh02}), large amounts of pristine gas can be funneled into the central regions of a dark matter halo through, e.g., the bars-within-bars instability \citep{shlosman89, shlosman90} or large-scale magnetic torques \citep{begelman23,hopkins23}. Depending on how efficiently angular momentum can be transported outward, a sufficiently large concentration of gas can be established on a short enough timescale to form a supermassive star that collapses due to a post-Newtonian instability (e.g., \citealt{hoyle63, fuller86, begelman10}), but the details of this process are almost certainly more complicated than the spherically symmetric ``Penston-Larson''-like models (e.g., \citealt{penston69, larson69, shu77, whitworth85, ogino99}) would suggest. An alternative is that the presence and distribution of angular momentum produces a more spatially distributed and centrifugally supported flow (i.e., a disk), resulting in a much smaller embryonic nuclear burning region that contains a correspondingly smaller fraction of the mass of the collapsing cloud. 

A variant on the latter scenario was proposed and investigated analytically by \citet{begelman06} (see also the numerical work by \citealt{choi13, shlosman16}), who emphasized that the bars-within-bars instability can yield a seed BH of only a few solar masses that is still surrounded by a large, gaseous envelope at the time of BH formation. The resulting structure --- a ``quasi-star'' --- is effectively a high-mass stellar envelope that is supported by black hole accretion (similar to, but distinct from, Thorne-Zytkow objects; \citealt{thorne75}, tentative observational evidence for which now exists; \citealt{levesque14}). \citet{begelman06, begelman08} suggested that such objects could exist and would promote extremely rapid BH growth until a limiting photospheric temperature was reached (analogous to the Hayashi limit; \citealt{hayashi61}), ultimately resulting in the ejection of the envelope \citep{begelman12} after most of the envelope mass was accreted.  This would yield a high-mass ($M_{\bullet} \gtrsim 10^{4-5}M_{\odot}$) BH seed at a redshift of $\gtrsim 10$. However, later investigations by \citet{ball11,ball12} found a lower limiting mass beyond which the BH mass could not grow, amounting to $M_{\bullet}/M_{\star} \simeq 0.0167$ with $M_{\star}$ the total quasi-star mass (see Section \ref{sec:quasi-stars-low} for a discussion of the different limits obtained when different inner boundary conditions are employed). If this limit is accurate, then it suggests that the quasi-star paradigm may not be capable of growing BHs to $\gtrsim$ IMBH masses.

Our purpose here is to reconsider the limiting mass of a BH growing inside a quasi-star, in light of a more thorough consideration of physical conditions in the quasi-star interior. In Section \ref{sec:quasi-stars-low}, we argue that the origin of the limit found first in \citet{ball11} and later elaborated in \citet{ball12} (with an error in \citealt{ball11} corrected) {is an artifact of an unphysical formulation of the inner boundary condition.} Instead, we suggest that when the BH grows to $\gtrsim 1\%$ of the total mass of the star, the innermost regions of the  envelope must resemble a hydrostatic atmosphere in a point-mass potential.  This implies that the energy sustaining the envelope must be generated well within this region, relatively close to the BH, and transported outward by strong convection.

Motivated by this argument, in Section \ref{sec:quasi-stars-moderate} we provide an alternative description for the inner regions of a quasi-star envelope in which the material is not in free fall (as assumed in \citealt{begelman08, ball11, ball12}), but {instead is drifting radially inward at a much lower speed, with convection transporting} the accretion energy outward at the maximum possible rate.  We show that these ``saturated convection'' solutions can be matched onto a surrounding, nearly adiabatic envelope within which the energy is transported in the presence of a weak entropy gradient, and that the {complete} solutions are determined solely by the ratio of the BH mass to quasi-star mass, $M_{\bullet}/M_{\star}$. These solutions permit a maximum BH mass $\sim 60\%$ (i.e., of order unity) of the total mass of the quasi-star, with most of the gas comprising the saturated convection region when this mass ratio is attained.

In Section \ref{sec:radii} we quantify the physical properties of a quasi-star (e.g., its radius and effective temperature) that result from this model, which can be determined by specifying the luminosity, which we equate to the Eddington limit of the total star, the BH mass ratio, and the total quasi-star mass. We summarize and discuss our results in Section \ref{sec:summary}.

\section{Quasi-stars in the low-black-hole-mass limit}
\label{sec:quasi-stars-low}
Expanding on the work of \citet{begelman06}, \citet{begelman08} analyzed quasi-stars in the limit that the accreting BH has a mass  $M_{\bullet} \ll M_{\star}$, where $M_{\star}$ is the total mass that includes the BH  and the surrounding envelope. In this case we can approximate the central region of the quasi-star as having $\rho , p , c_{\rm s} \simeq$ {\it const.}, where $\rho$, $p$ and $c_{\rm s} = \sqrt{\gamma p/\rho}$ are the density, pressure, and sound speed near the center of the quasi-star; these conditions should hold until we approach the sphere of influence of the central BH. \citet{begelman08} argued that these conditions are qualitatively similar to those that describe Bondi accretion \citep{bondi52}, where gas with a constant sound speed at asymptotically large distances from a point mass is accreted in a spherically symmetric manner. A supermassive star is radiation-pressure dominated and convective throughout the majority of its interior by both mass and radius (e.g., \citealt{loeb94, hansen04}), and hence the quasi-star envelope is well-approximated by a $\gamma = 4/3$ polytrope. \citet{begelman08} therefore concluded that a quasi-star should resemble a $\gamma = 4/3$ polytrope down to a radius
\begin{equation}
r_{\rm B} \simeq N\frac{2GM_{\bullet}}{c_{\rm s}^2}, \label{bondicond}
\end{equation}
which is the sphere of influence of the BH and is effectively synonymous with the Bondi radius. Here $N$ is a number of order unity that can be tuned to any arbitrary value and reflects our ignorance of the details of the transition between these two types of flow (i.e., hydrostatic and effectively free fall). In the \cite{begelman08} treatment, the sound speed appearing on the right-hand side of this expression is approximated as that of a $\gamma = 4/3$ polytrope --- constant in the limit approaching the center of the star.

\citet{ball11} reanalyzed quasi-stars using the Cambridge STARS stellar evolution code (e.g., \citealt{eggleton71, eldridge04}), and pointed out that as the BH mass grows, it may no longer be accurate to ignore the gravitational influence of the BH on the quasi-star envelope. They included the BH mass in computing the structure of the envelope (calculated numerically with the STARS code) and continued to employ Equation \eqref{bondicond} as the boundary condition in the interior, but with $c_{\rm s}$ now accounting for the effects of the BH on the envelope structure. \citet{ball11} found that there was a critical BH mass, of order $M_{\bullet} \simeq 0.1 M_{\star}$, above which they could not find a solution that satisfied the stellar structure equations and Equation \eqref{bondicond}. \citet{ball12} corrected the boundary condition used in \citet{ball11} to account for the gas mass contained within the Bondi flow (the density profile of which satisfies $\rho \propto r^{-3/2}$; in other words, they replaced $M_{\bullet}$ appearing in Equation \ref{bondicond} with the total mass contained within the Bondi radius), and found that --- when the outer envelope is approximated by a $\gamma = 4/3$ polytrope and $N = 1$ in Equation \eqref{bondicond} --- the black hole mass could not exceed $\sim 0.0167$ of the total mass of the star, i.e., $M_{\bullet}/M_{\star} \le 0.017$. \citet{ball11, ball12} drew a parallel between this limit and the  Sch\"onberg-Chandrasekhar \citep{SC42} limit on the masses of inert cores inside stellar envelopes, but noted that the physical origin for the limit in the quasi-star case remained unclear. If true, this would limit the validity of the quasi-star model for BH growth to very early stages when $M_\bullet \ll M_*$. 

One can easily reproduce this limit by assuming that the envelope structure is given by a $4/3$-polytrope with $p = K\rho^{4/3}$. The equation of hydrostatic equilibrium is
\begin{equation}
    \tilde{K}\frac{d}{d\xi}\left[\left(\frac{1}{\xi^2}\frac{dm}{d\xi}\right)^{1/3}\right] = -\frac{m}{\xi^2}, \label{dimle}
\end{equation}
where
\begin{equation}
    M = M_{\rm B}m(\xi), \quad \xi = \frac{r}{r_{\rm i}} = \frac{r}{r_{\rm B}}, \quad \tilde{K} = \frac{4K M_{\rm B}^{-2/3}}{G\left(4\pi\right)^{1/3}}.
\end{equation}
Here $M_{\rm B}$ is the mass contained within the Bondi radius, including both the BH mass and the gas mass within that region, and $M$ is the mass contained within $\xi$ that is related to the density via
\begin{equation}
    \rho = \frac{M_{\rm B}}{4\pi r_{\rm B}^3}\frac{1}{\xi^2}\frac{dm}{d\xi}.
\end{equation}
It follows that $m(1) = 1$ to conserve the mass in going from the inner, freefalling region to the hydrostatic envelope. Using these definitions in the expression for the Bondi radius \eqref{bondicond} but replacing  $M_{\bullet}$ by $M_{\rm B}$  then shows that
\begin{equation}
    \frac{dm}{d\xi}\bigg{|}_{\xi = 1} = \left(\frac{6N}{\tilde{K}}\right)^3.
\end{equation}
$\tilde{K}$ is not immediately constrained by these relations, but if we assume that the polytropic/adiabiatic envelope extends to very near the photospheric radius of the star, then the total mass of the quasi-star --- which is determined by where $m'(\xi) \simeq 0$ --- is a function only of $\tilde{K}$ (for a given $N$) and this closes the system. We can then simply integrate Equation \eqref{dimle} as we vary $\tilde{K}$ to determine the solutions as a function of the ratio of $M_{\rm B}/M_{\star}$. We can also relate $M_{\rm B}$ to the black hole mass, $M_{\bullet}$, if we let the density profile interior to $r_{\rm B}$ be $\rho = \rho_{\rm B}\left(r/r_{\rm B}\right)^{-3/2}$ with $\rho_{\rm B}$ given by the envelope density at $r_{\rm B}$, which yields
\begin{equation}
    \frac{M_{\bullet}}{M_{\star}} = \frac{M_{\rm B}}{M_{\star}}\left(1-\frac{2}{3}\left(\frac{6N}{\tilde{K}}\right)^3\right).
\end{equation}
Because we require $M_{\bullet}/M_{\star} > 0$, this sets the following limit on $\tilde{K}$:
\begin{equation}
    \tilde{K} > 6N\left(\frac{2}{3}\right)^{1/3} \simeq 5.24 N.
\end{equation}

\begin{figure}
    \includegraphics[width=0.475\textwidth]{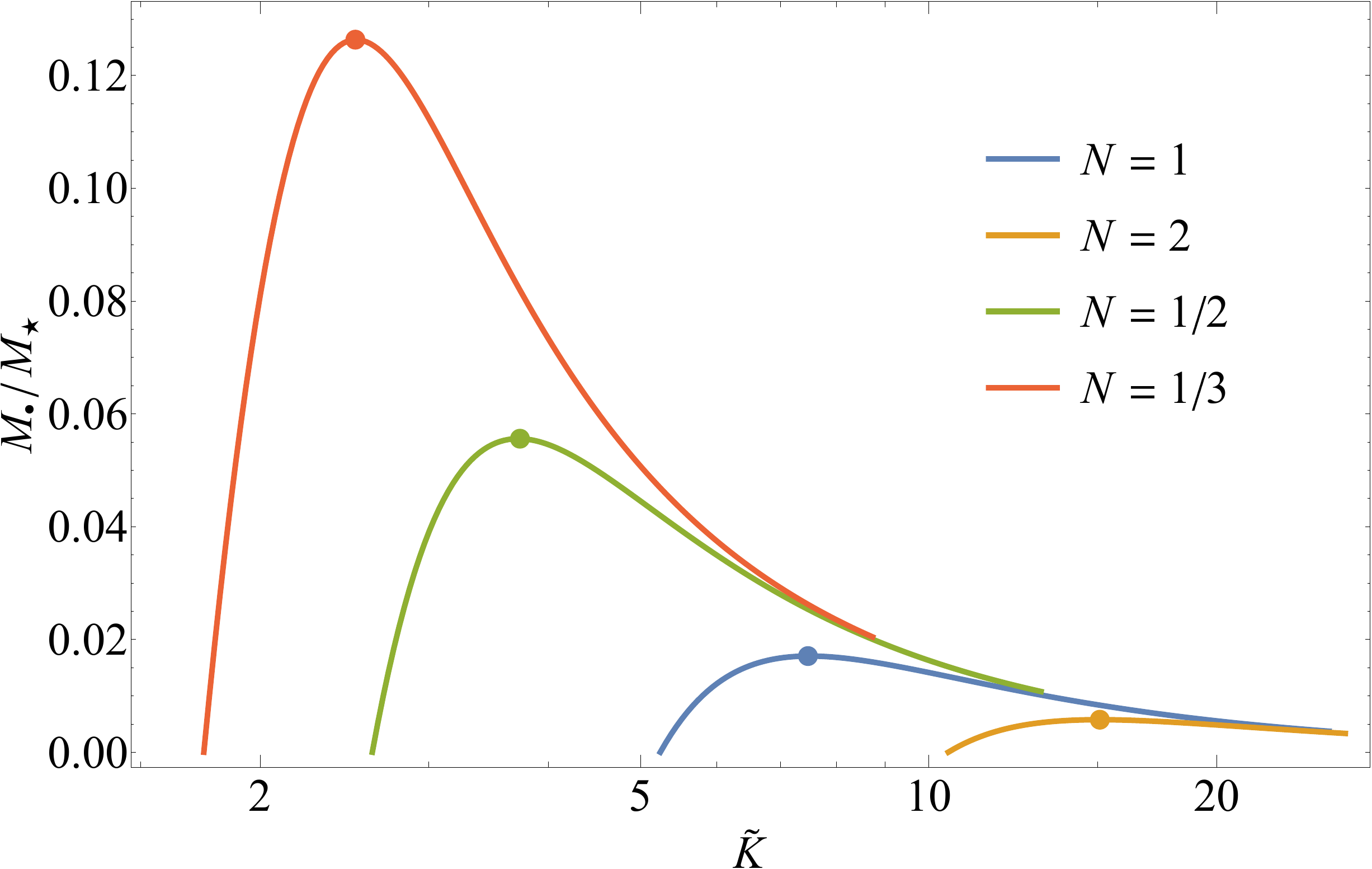}
    \caption{The ratio of the BH mass to the total quasi-star mass as a function of the dimensionless entropy, $\tilde{K}$, that enters into the Lane-Emden equation. The points denote the maximum possible ratio that can be achieved, where the different curves correspond to the $N$ in the legend, which is the ratio of the inner radius to the Bondi radius.}
    \label{fig:ballplot}
\end{figure}

Figure \ref{fig:ballplot} shows the ratio of the BH mass to the quasi-star mass as a function of $\tilde{K}$, with the blue curve appropriate to $N = 1$ --- the case analyzed by \citet{ball11, ball12}. Since $m'(1) \propto 1/\tilde{K}^3$, small mass ratios physically correspond to large values of $\tilde{K}$, as for these solutions the outer radius is much greater than $r_{\rm B}$. We see that the blue curve, which corresponds to $N = 1$, reaches a maximum value of $\sim 0.017$ and subsequently declines. There are no solutions with $M_\bullet / M_* \gtrsim 2\%$ for this value of $N$.  Since we generally expect the mass ratio to increase with time, the system should evolve along the curve from right to left with BH growth terminating at the local maximum, marked by a filled circle.  

However, there is no reason to enforce $N = 1$, and we can vary this parameter to understand its influence on the maximum BH mass. The three other curves in Figure \ref{fig:ballplot} show that the BH mass is an extremely sensitive function of $N$: increasing $N$ by a factor of 2 (yellow curve) reduces the maximum mass to $\sim 0.0058$, while setting $N = 1/3$ increases the maximum mass to $\sim 13\%$ $M_{\star}$ --- nearly a factor of 10 larger than that obtained with $N = 1$. By further reducing the value of $N$, we can obtain black hole masses that are order-unity fractions of the total quasi-star mass. This ``limit'' is therefore unphysical, given its strong dependence on an unknown and entirely tunable parameter. 

\begin{figure*}
    \includegraphics[width=0.490\textwidth]{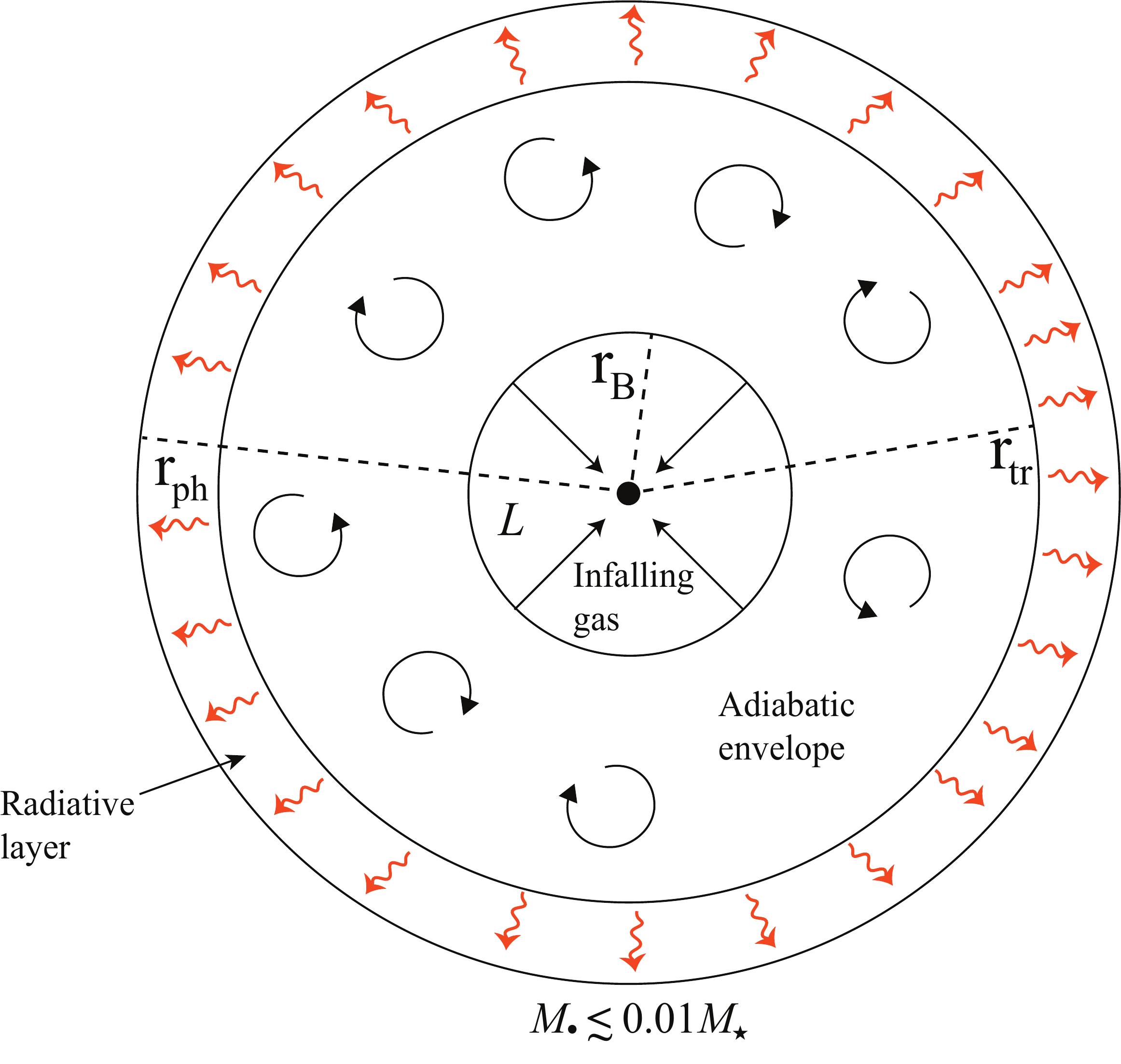}
    \includegraphics[width=0.51\textwidth]{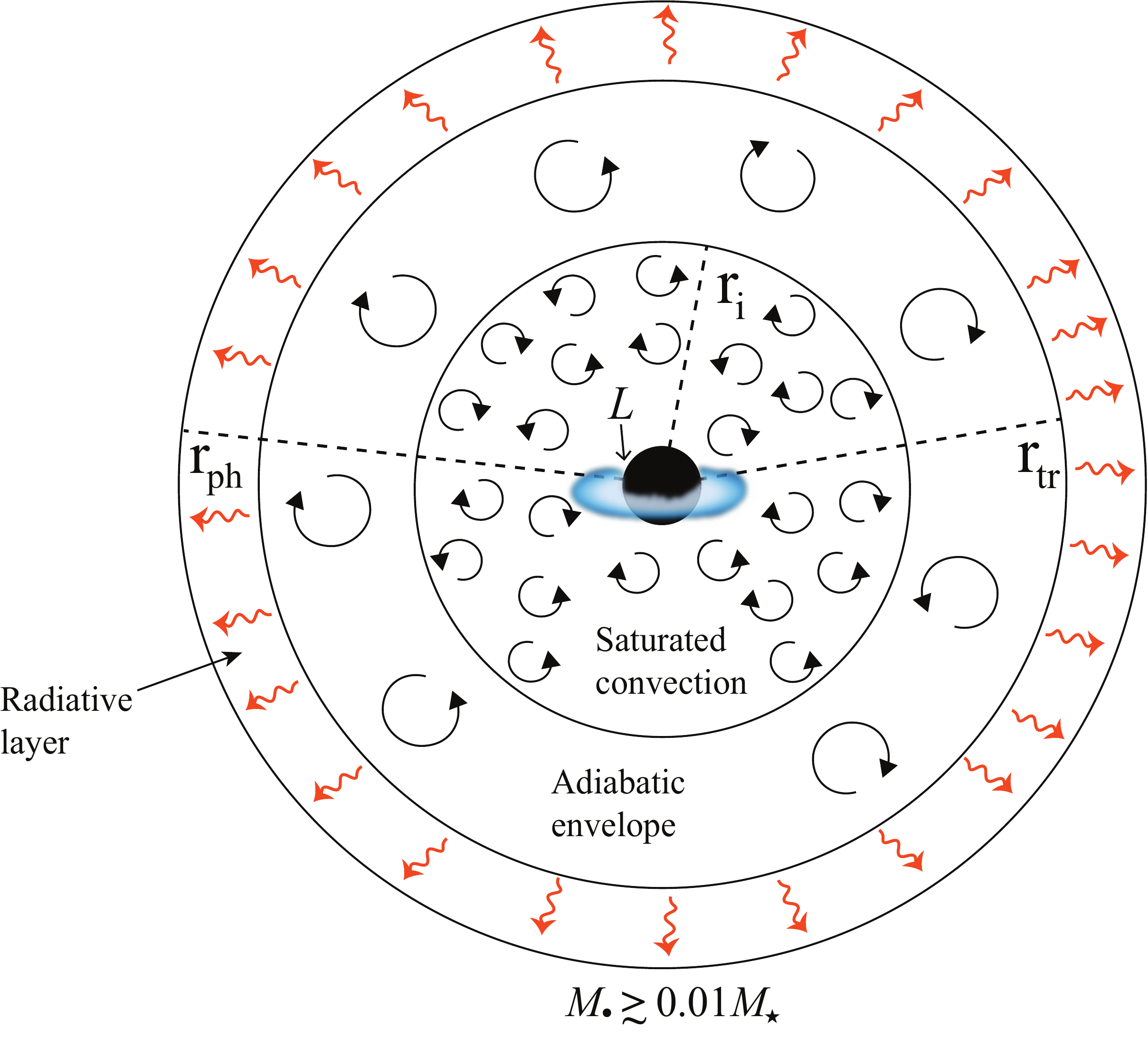}
      \caption{{A cartoon illustrating two quasi-star structures: the left applies when the black hole mass is a small fraction of the total quasi-star mass, and consists of an inner region within which the gas is infalling onto the black hole; this setup was adopted by \citet{begelman08, ball11, ball12}, and only exists for $M_{\bullet}/M_{\star} \lesssim 0.01$. The radius $r_{\rm B}$ is the Bondi radius, which separates the infalling region and the outer, $\sim$ adiabatic and convective envelope, which transitions to a radiative layer at $r_{\rm tr}$ (which itself ends in the photosphere at radius $r_{\rm ph}$). In this case, the energy (luminosity $L$) is liberated as material falls in within $r_{\rm B}$ and accretes, and is carried out by convection at a rate that is set by conditions near the Bondi radius (see the discussion in Section 2.1 of \citealt{begelman08}). The picture on the right --- which is the focus of this paper --- similarly consists of three connected regions, but the Bondi-like flow at small radii is replaced by one within which convection transports the energy outward at the local and maximum possible rate, with a corresponding luminosity $L \simeq 4\pi r^2 p c_{\rm s}$. In this picture, most of the energy is generated very near the black hole through an accretion disc (cartoonishly depicted in blue), implying that the hydrostatic region extends much deeper into the interior of the star. Note that neither of these depictions reflects the true disparity in scales between the black hole horizon and the photospheric radius.}}
      \label{fig:schematic}
\end{figure*}

In exploring the low-mass regime, \citet{begelman08} proposed that the accretion rate onto the black hole was suppressed below the Bondi rate, to allow convection near $r_{\rm B}$ to carry the liberated energy. As the BH  mass grows and starts to more significantly affect the envelope around it, the increasing central sound speed implies that convection should be able to more readily transport the accretion energy, allowing more gas to reach the BH. Motivated by this observation, in the next section we propose a model in which the inner regions of the star transport accretion energy at the theoretical maximum rate achievable by convection. We show that these ``saturated-convection'' solutions can smoothly join onto an exterior, $\gamma = 4/3$, polytropic envelope for black holes with masses in excess of those determined by imposing Equation \eqref{bondicond}, thus enabling the mass of the black hole to grow into the regime where $M_{\bullet}/M_{\star} \sim$ {\it few} $\times 0.1$ on cosmologically short timescales. {Figure \ref{fig:schematic} gives an illustration of the quasi-star model as put forth by \citet{begelman08, ball11, ball12} (left) and the new model described in this paper (right). Note that in these previous investigations, the gas is not in pure freefall within $r_{\rm B}$, but is infalling at a rate that is reduced to allow the accretion energy to be transported outward; see Section 2.1 of \citet{begelman08}. }

\section{Quasi-stars in the moderate-black-hole-mass limit}
\label{sec:quasi-stars-moderate}
\subsection{Inner, saturated-convection solutions}
\label{sec:inner}
Let us assume that the inner region of the quasi-star (i.e., where the BH dominates the dynamics) behaves in such a way that the majority of the energy is produced near the BH and strong convection is the dominant mode of energy transport. We expect the convective transport speed to saturate at close to the sound speed (e.g., \citealt{begelman08}), since supersonic convection should quickly lead to shock formation and the dissipation of energy. In this case the convective energy flux is
\begin{equation}
F_{\rm con} \simeq \beta p c_{\rm s},
\end{equation}
with $\beta \lesssim 1$ an efficiency parameter, and the convective luminosity is given by
\begin{equation}
L \simeq 4\pi r^2 F_{\rm con} = 4\pi r^2 \beta p c_{\rm s}.
\end{equation}
Under the assumption that most of the energy is generated near the black hole, $L$ is constant, and setting $\gamma = 4/3$ for the adiabatic index gives the following relationship between the pressure and density that acts effectively as an equation of state:
\begin{equation}
p = \left(\frac{\sqrt{3}L}{8\pi\beta}\right)^{2/3}r^{-4/3}\rho^{1/3}. \label{peos}
\end{equation}

Suppose there is a radius $r_{\rm i}$ that divides the inner, saturated convection region and the outer, polytropic envelope that is energetically sustained by approximately adiabatic convection. The mass contained within $r_{\rm i}$ is $M_{\rm i}$, and we define scaled radial and mass coordinates by $\xi = r/r_{\rm i}$ and $m_{\rm i} = M(r)/M_{\rm i}$. Using Equation \eqref{peos} in the equation of hydrostatic equilibrium with the usual relation for mass conservation then gives
\begin{equation}
    K_{\rm i}\frac{d}{d\xi}\left[\xi^{-2}\left(\frac{dm_{\rm i}}{d\xi}\right)^{1/3}\right] = -\frac{m_{\rm i}}{\xi^4}\frac{dm_{\rm i}}{d\xi}, \label{lanein}
\end{equation}
where
\begin{equation}
    K_{\rm i} = \left(\frac{L\sqrt{3}}{2\beta}\right)^{2/3}\left(\frac{r_{\rm i}}{M_{\rm i}}\right)^{5/3}\frac{1}{G}. \label{Kieq}
\end{equation}
This equation holds for all $\xi \le 1$, subject to the boundary condition that $m_{\rm i}(1) = 1$.

While the derivative of the mass at $\xi = 1$ does not appear to be immediately constrained, requiring that the solutions extend to $\xi \rightarrow 0$ allows us to relate $m'_{\rm i}(1)$ to $K_{\rm i}$. Specifically, if we redefine the variables $\xi$ and $m_{\rm i}$ by $\tilde{\xi} = \xi/\alpha$ and $\tilde{m}_{\rm i} = {m}_{\rm i}/m_{\bullet}$, where $m_{\bullet}$ is the value of $m_{\rm i}$ in the limit that $\xi \rightarrow 0$ and $\alpha = m_{\bullet}K_{\rm i}^{-3/5}$, then the above equation becomes
\begin{equation}
    \frac{d}{d\tilde{\xi}}\left[\tilde{\xi}^{-2}\left(\frac{d\tilde{m}_{\rm i}}{d\tilde{\xi}}\right)^{1/3}\right] = -\frac{\tilde{m}_{\rm i}}{\tilde{\xi}^4}\frac{d\tilde{m}_{\rm i}}{d\tilde{\xi}}. \label{mitilde}
\end{equation}
We can now expand this solution about $\tilde{\xi} = 0$, which gives
\begin{equation}
\tilde{m}_{\rm i}(\tilde{\xi} \simeq 0) \simeq 1+\frac{3\sqrt{3}}{5\sqrt{2}}\tilde{\xi}^{5/2}, \label{mapp}
\end{equation}
and use this to integrate Equation \eqref{mitilde} outward, i.e., $\tilde{m}_{\rm i}$ is a unique function of $\tilde{\xi}$ that can be determined numerically. 

To satisfy the boundary condition $m_{\rm i}(1) = 1$, we require
\begin{equation}
    m_{\bullet}\tilde{m}_{\rm i}\left(\tilde \xi = \frac{K_{\rm i}^{3/5}}{m_{\bullet}}\right) = 1. \label{Kiofmbullet}
\end{equation}
For a given $K_{\rm i}$, this expression will be satisfied for some $m_{\bullet}$, thus relating these two quantities. We then have
\begin{equation}
    m_{\rm i}'(\xi = 1) = K_{\rm i}^{3/5}\tilde{m}_{\rm i}'\left(\tilde \xi = \frac{K_{\rm i}^{3/5}}{m_{\bullet}}\right),
\end{equation}
where $m_{\rm i}' = dm_{\rm i}/d\xi$ and $\tilde{m}_{\rm i}' = d\tilde{m}_{\rm i}/d\tilde{\xi}$. Since $m_{\bullet}$ is a function of $K_{\rm i}$, this shows that $m'_{\rm i}(1)$ is also a function of only $K_{\rm i}$, i.e., $K_{\rm i}$ (or the mass of the BH, $m_{\bullet}$) is the only free parameter in determining the saturated convection solutions.

\begin{figure}
    \includegraphics[width=0.475\textwidth]{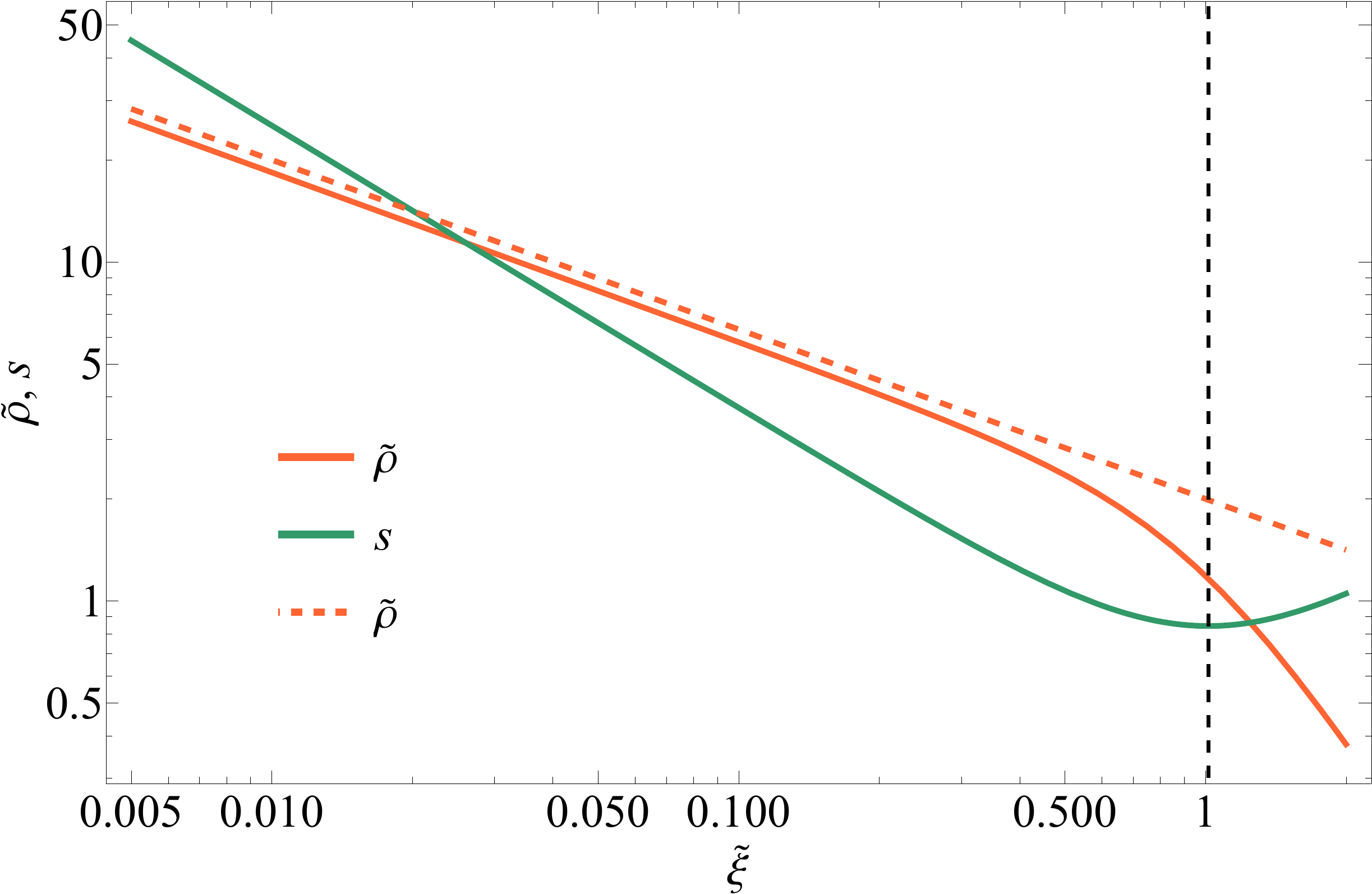}
 \caption{The dimensionless density $\tilde{\rho}$ and entropy $s$ as functions of $\tilde{\xi}$. The vertical, dashed line shows where the gradient in the entropy equals zero and is the maximum $\tilde{\xi}$, equal to $\tilde{\xi} \simeq 1.015$, to which these solutions can physically extend.}
    \label{fig:rho_of_xitilde}
\end{figure}

We also note that there is a maximum value of $K_{\rm i}$ (or a minimum value of $m_{\bullet}$) that these solutions permit: Figure \ref{fig:rho_of_xitilde} shows the density (orange) $\tilde{\rho} = \tilde{m}'/\tilde{\xi}^2$ and the entropy (green) $s = \tilde{p}/\tilde{\rho}^{4/3}$ (where $\tilde{p} = \tilde{\xi}^{-4/3}\tilde{\rho}^{1/3}$) as functions of $\tilde{\xi}$. Because we require this region to be strongly convective, the entropy must decrease outward, which is only satisfied for $\tilde{\xi} \le 1.015$; at $\tilde{\xi} \simeq 1.015$ the entropy reaches a relative minimum and thereafter increases. Setting $\xi = 1$ and $\tilde{\xi} = 1.015$ in $\tilde{\xi} = \xi/\alpha$, using $\alpha = m_{\bullet}K_{\rm i}^{-3/5}$ and the relationship between $m_{\bullet}$ and $K_{\rm i}$, we find that $K_{\rm i} \le 0.46554$ and $m_{\bullet} > 0.6321$. For this value of $K_{\rm i}$, then, the gas mass in the saturated convection region contributes $\sim 37\%$ of the total mass, while the BH mass constitutes the remainder. 

In the next section we describe how to construct the outer, polytropic solutions. Before doing so, however, we note that the saturated-convection solutions satisfy $\rho \propto \xi^{-1/2}$ and $p \propto \xi^{-3/2}$ at $\xi \ll 1$. These are the same density and pressure scalings as the convection-dominated accretion flow (CDAF) solutions described in \citet{quataert00}. One can therefore consider these solutions as spherically averaged, self-gravitating CDAFs, which are ultimately truncated at a radius where self-gravity results in a more rapid decline in the density and pressure than can self-consistently support the outward transport of energy via convection.

\subsection{Outer, polytropic envelope}
\label{sec:outer}
Outside of $r_{\rm i}$ is the polytropic envelope that contains most of the mass at early times. This envelope should also be dominated by radiation pressure, hence the pressure and density are related via $p = K_{\rm o}\rho^{4/3}$. The outer envelope therefore satisfies the Lane-Emden equation with $\gamma = 4/3$, which when written in terms of the same normalized variables (i.e., letting $m_{\rm o} \equiv M/M_{\rm i}$ and $\xi \equiv r/r_{\rm i}$) is
\begin{equation}
    \tilde{K}_{\rm o}\frac{d}{d\xi}\left[\left(\frac{1}{\xi^2}\frac{dm_{\rm o}}{d\xi}\right)^{1/3}\right] = -\frac{m_{\rm o}}{\xi^2}, \label{laneout}
\end{equation}
where
\begin{equation}
    \tilde{K}_{\rm o} = \frac{4K_{\rm o}}{GM_{\rm i}^{2/3}\left(4\pi\right)^{1/3}}.
\end{equation} 

In crossing $r_{\rm i}$, the mass, pressure, and temperature must be continuous; the latter is required to prevent an infinite radiative and diffusive flux at this location. Mass continuity at $r_{\rm i}$ gives $m_{\rm o}(1) = 1$. If the star is composed purely of ideal gas and radiation and there is no change in the mean molecular weight across $r_{\rm i}$, then the continuity of the pressure and temperature means that the density is also continuous. Taken together, the continuity of these quantities implies that the boundary conditions on the previous functions become, in addition to $m_{\rm i}(1) = m_{\rm o}(1) = 1$,
\begin{equation}
    m_{\rm o}'(1) = m_{\rm i}'(1), \quad \tilde{K}_{\rm o} = 4K_{\rm i}m_{\rm i}'(1)^{-1},
\end{equation}
where primes denote derivatives with respect to $\xi$. Since $m_{\rm i}'(1)$ depends only on $K_{\rm i}$, the polytropic solutions can be integrated from $\xi = 1$ outward exclusively as a function of this parameter.

The solution to the Lane-Emden equation can extend to a radius at which the density equals zero, but we expect the envelope to transition to a radiative zone --- where convection is no longer efficient enough to transport the energy --- prior to this location. We can estimate where this transition occurs by noting that the maximum convective flux is still $L_{\rm con, max} = 4\pi r^2\beta p c_{\rm s}$, where the pressure and sound speed can be deduced from the polytropic profiles. Once this luminosity equals the total luminosity, $L$, radiative diffusion must become the dominant mode of energy transport. We therefore expect the convective envelope to extend to a radius, $\xi_{\rm r}$, where
\begin{equation}
    L_{\rm con, max} = 4\pi r^2\beta p c_{\rm s} = L,
\end{equation}
which, upon using our previously defined variables, gives the condition
\begin{equation}
    \frac{1}{\xi_{\rm r}}\left(\frac{m_{\rm o}'(\xi_{\rm r})}{m_{\rm o}'(1)}\right)^{3/2} = 1.
\end{equation}
The $K_{\rm i}$-dependent value of $\xi_{\rm r}$ at which this equality is satisfied delimits the radius at which the polytropic envelope ends and the radiative layer --- which ultimately terminates in the photosphere --- begins. 

The total mass of the quasi-star, $M_{\star}$, has a contribution from the radiative layer. However, if the radiative layer is thin, then the mass at $\xi_{\rm r}$ approximately equals the total mass. In this approximation, the relative BH mass $M_{\bullet}/M_{\rm i}$ and the total quasi-star mass relative to $M_{\rm i}$, $M_{\star}/M_{\rm i}$, are functions only of $K_{\rm i}$. It therefore follows that the ratio $M_{\bullet}/M_{\star}$ is determined purely by the variable $K_{\rm i}$, and by varying this quantity (from Equation \ref{Kieq}, $K_{\rm i}$ relates the radius $r_{\rm i}$ to the luminosity $L$ and mass $M_{\rm i}$), we can assess the maximum BH mass (relative to the quasi-star mass) that is achievable without any reference to the other physical quantities. In this way the solutions are analogous to those that impose the inner boundary condition of the Bondi radius, as discussed in Section \ref{sec:quasi-stars-low}, but here there are no free parameters. 

\begin{figure*}[htbp] 
   \centering
   \includegraphics[width=0.495\textwidth]{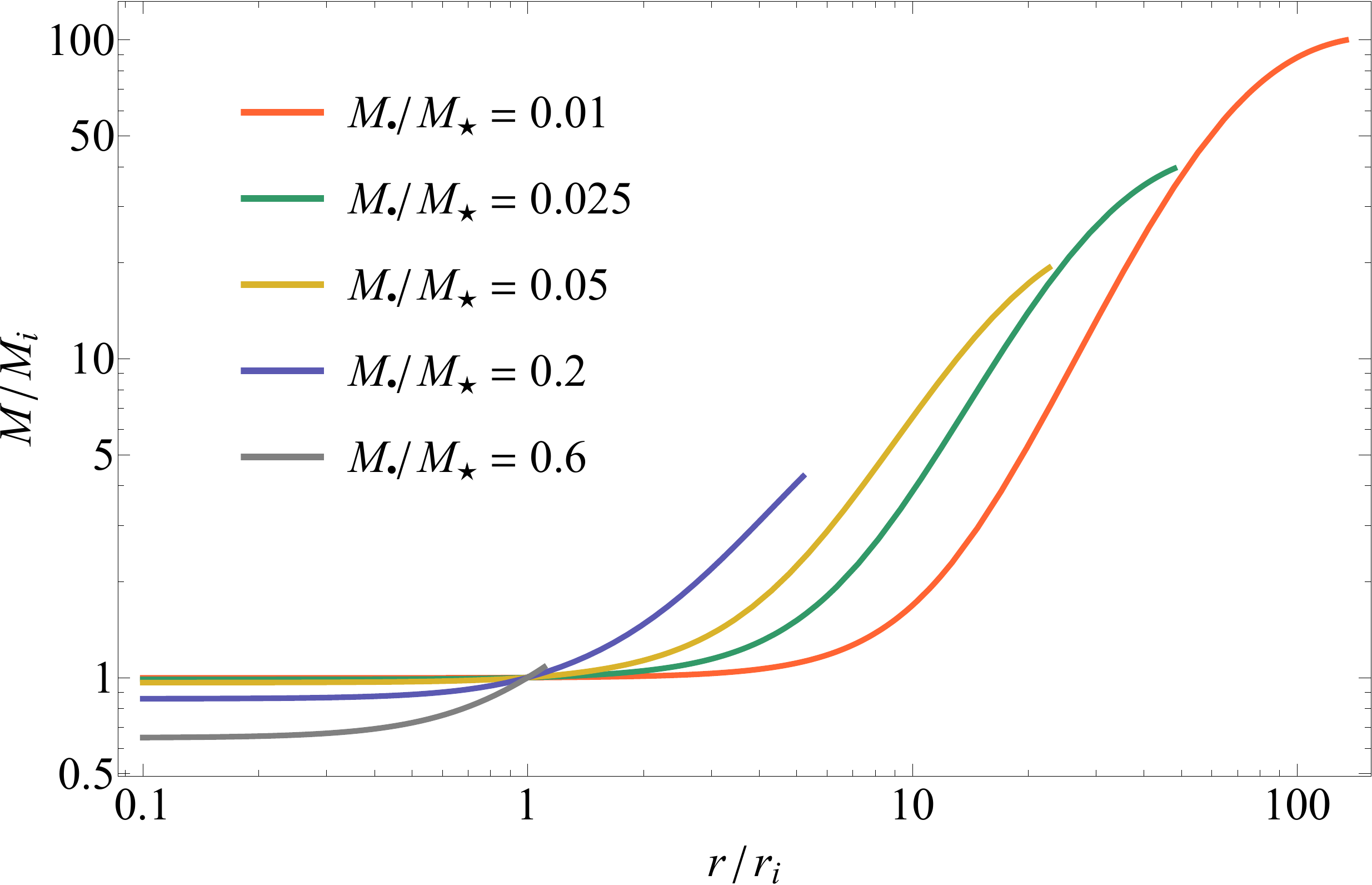} 
   \includegraphics[width=0.495\textwidth]{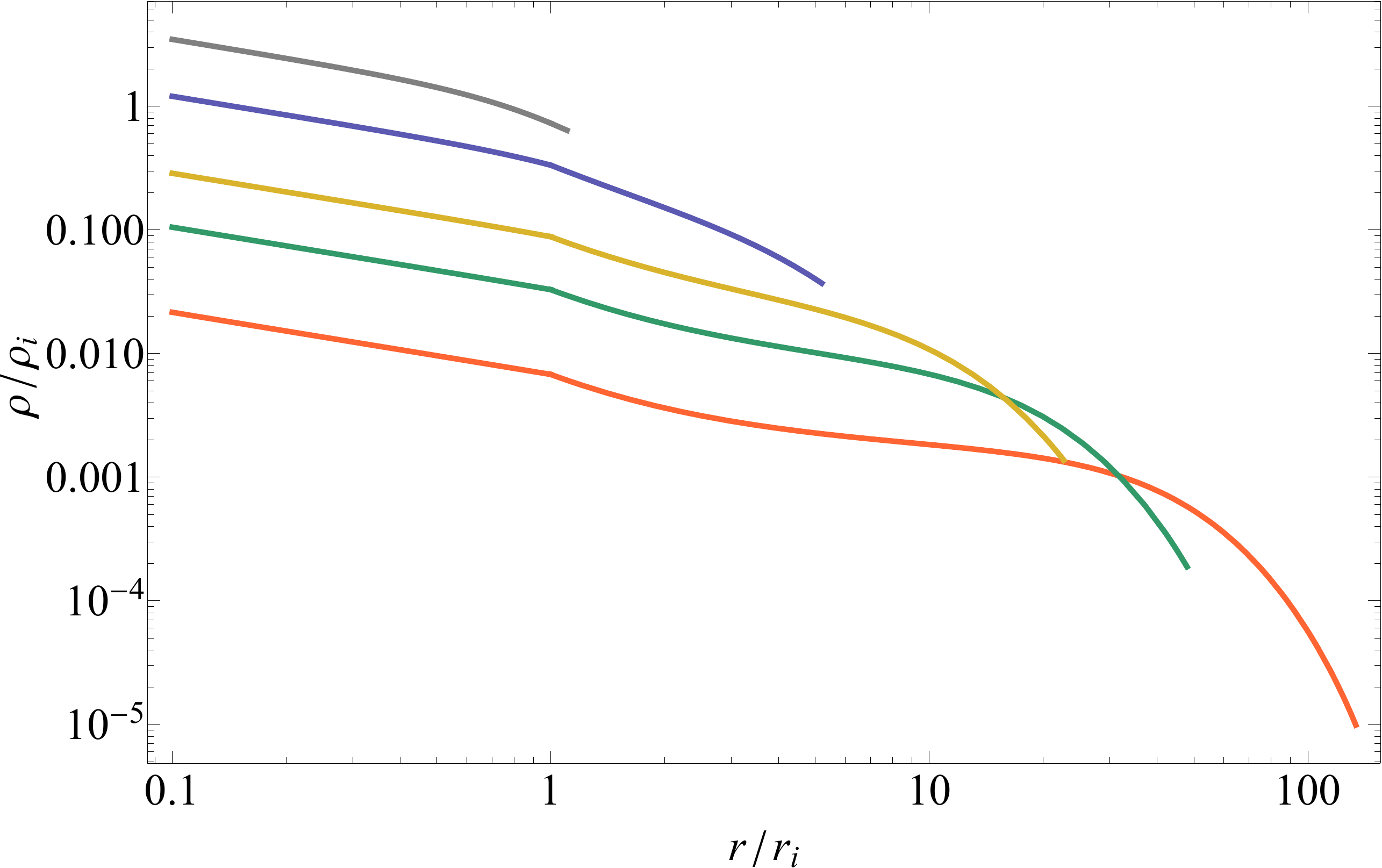} 
   \caption{Left: the normalized mass enclosed within radius $r$, $M/M_{\rm i}$, for the BH mass ratios shown in the legend. The solutions are continuous across $r = r_{\rm i}$, interior (exterior) to which the solutions are governed by the saturated convection (polytropic) equation of state. As the mass ratio increases, the outer radius shrinks, and most of the mass is contained interior to $r_{\rm i}$. Right: the normalized density as a function of radius for the same mass ratios shown in the left panel, where $\rho_{\rm i} = M_{\rm i}/(4\pi r_{\rm i}^3)$.  The density is continuous across $r_{\rm i}$, although there is a discontinuity in the density gradient that is more pronounced as the mass ratio decreases.}
   \label{fig:mass_of_r}
\end{figure*}

Figure \ref{fig:mass_of_r} illustrates the mass contained within radius $r$ (left) and the density (right) as a function of radius normalized by $r_{\rm i}$, where the scale density is $\rho_{\rm i} = M_{\rm i}/(4\pi r_{\rm i}^3)$. The different curves correspond to the mass ratios shown in the legend (the coloration of the curves in the right panel correlates with the legend in the left panel). Both the mass and density are continuous across $r_{\rm i}$, which is a consistency check on the boundary conditions that we impose at this radius. For small mass ratios the outer radius of the star satisfies $\xi_{\rm r} \gg 1$, meaning that the envelope is highly extended, and most of the mass is contained in the polytropic region ($r > r_{\rm i}$). As the mass ratio increases, the radius of the polytropic envelope is pushed to smaller radii (in units of $r_{\rm i}$), and more of the mass is contained within $r_{\rm i}$. For $M_{\bullet}/M_{\star} = 0.6$, the polytropic envelope is almost nonexistent ($\xi_{\rm r} \simeq 1.1$), with most of the remaining $\sim 40\%$ of the mass contained within the gas that comprises the saturated convection region.

\begin{figure}
\includegraphics[width=0.475\textwidth]{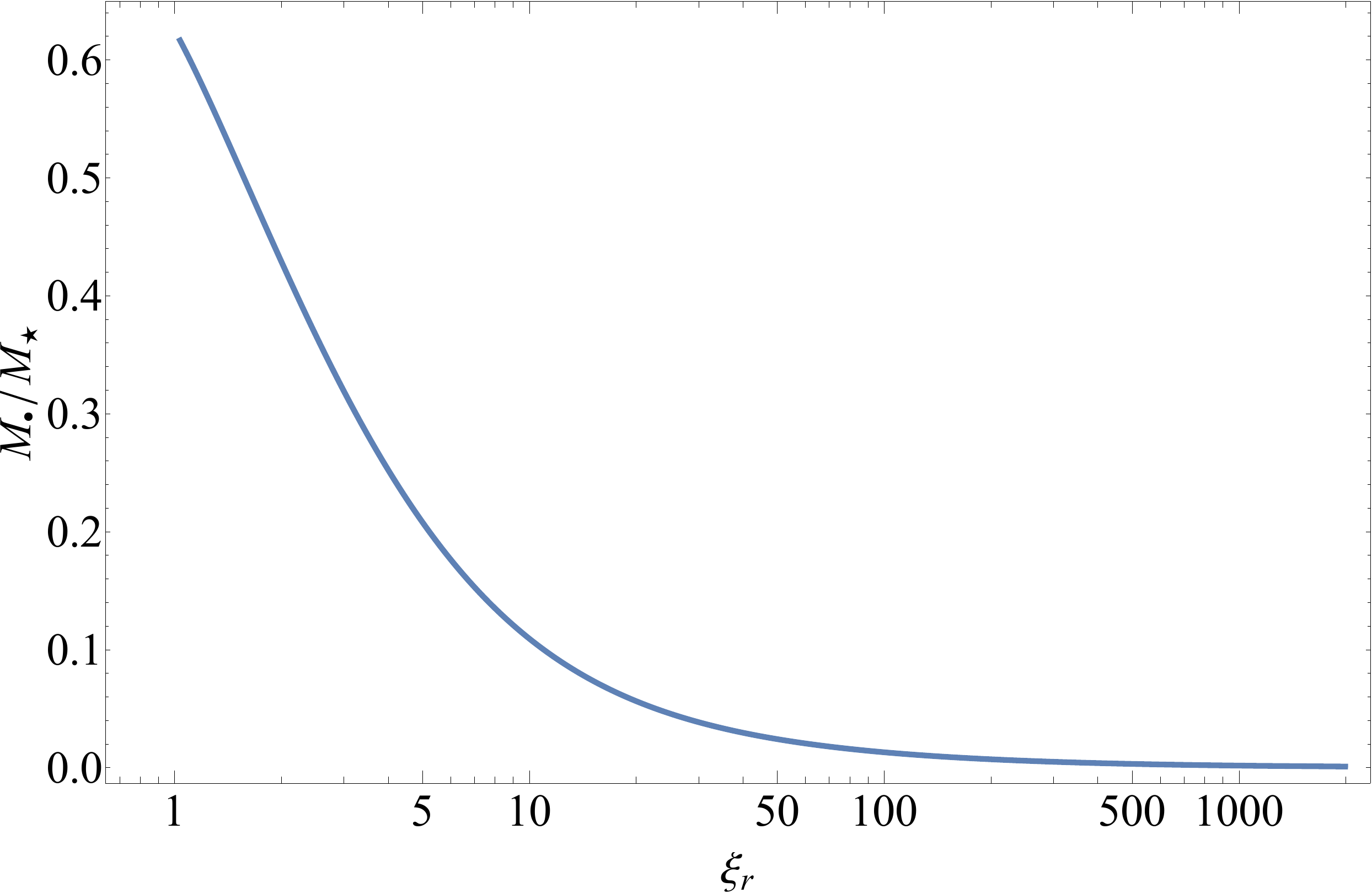}
    \caption{The ratio of the BH mass to the total quasi-star mass, $M_{\bullet}/M_{\star}$, obtained from the saturated-convection solutions, as a function of the dimensionless radius that divides the convective and radiative regions, $\xi_{\rm r}$. As the BH mass grows, the saturated convection region comprises more of the quasi-star, reaching $\xi_{\rm r} \simeq 1$ (and disappearing almost entirely) when $M_{\bullet}/M_{\star} \simeq 0.618$.}
    \label{fig:mbullet_saturated}
\end{figure}

Figure \ref{fig:mbullet_saturated} shows the BH mass ratio $M_{\bullet}/M_{\star}$ as a function of $\xi_{\rm r}$, both of which are functions of $K_{\rm i}$. At small mass ratios, the transition to the radiative layer {lies far outside the saturated convection zone, $\xi_r \gg 1$}, implying that the saturated convection region comprises a very small fraction of the quasi-star by volume (and mass). As the mass ratio grows, the polytropic region shrinks in size, and when $M_{\bullet}/M_{\star} \simeq 0.618$, $K_{\rm i} = 0.4655$ --- the maximum value of $K_{\rm i}$ permitted by the solutions, as described in Section \ref{sec:inner} --- and the {polytropic zone nearly vanishes ($\xi_{\rm r} \simeq 1.02$). }  

This model therefore predicts a maximum BH mass of $M_{\bullet}/M_{\star} \simeq 0.618$, with $\sim 36.8$\% of the mass contained within the gas that comprises the saturated convection region, and the remaining $\sim 2.3$\% within the polytropic envelope. Unlike the solutions that match onto freefall in the inner regions, the saturated convection inner region --- which allows the accretion energy to be liberated near the BH and transported outward at the maximum achievable rate to join onto the polytropic envelope ---  permits order-unity ratios of the BH mass to the total quasi-star mass. In the next section we relate these self-similar solutions to physical solutions, and we describe the growth of the BH by demanding that the luminosity equal the Eddington limit for the total mass. 

\section{Quasi-star Physical Properties and Black Hole Growth}
\label{sec:radii}
The solutions described in the previous section are self-similar, meaning that the physical scales (e.g., the radius of the quasi-star) need not be specified in order to construct the solutions that depend only on the ratio of the BH to quasi-star mass. However, we break self-similarity once we specify the luminosity, which we expect to be close to the Eddington limit of the total star. The reason for this is that $L_{\rm rad} \simeq L$ once radiative diffusion takes over as the dominant mode of energy transport (at $\sim \xi_{\rm r}$), but since the star is radiation-pressure dominated and in hydrostatic equilibrium, it follows that $L_{\rm rad} \simeq L_{\rm Edd} = 4\pi GM_{\star} c/\kappa$ with $\kappa$ the opacity at the beginning of the radiative layer. If the temperature at the base of the radiative layer is large enough and the density low enough, we expect $\kappa \simeq \kappa_{\rm es} \simeq 0.34$ cm$^2$ g$^{-1}$. 

Assuming that BH accretion supplies the energy supporting the envelope, we have
\begin{equation}
    \frac{4\pi G M_{\star} c}{\kappa} = \eta c^2\frac{dM_{\bullet}}{dt}, \label{mdoteq}
\end{equation}
where we will make the usual assumption that $\eta = 0.1$. If the mass of the quasi-star grows linearly with time as $M_{\star}(t) = M_{0}+\dot{M}_{\star}t$ and the initial BH mass is $\ll M_{0}$, then Equation \eqref{mdoteq} can be integrated to yield
\begin{equation}
    \frac{M_{\bullet}}{M_{\star}} = \frac{4\pi G}{\eta\kappa c}t\frac{1+\frac{\dot{M}_{\star}t}{2M_0}}{1+\frac{\dot{M}_{\star}t}{M_0}}.
\end{equation}
In the limits of $\dot{M}_{\star} \simeq 0$ and $\dot{M}_{\star} t \gg M_0$, the mass ratio increases linearly with time, with the growth rate differing between the two limits by a factor of 2. The time taken for the BH mass ratio to reach $\sim 0.6$ --- the maximum value achievable via these solutions --- is thus
\begin{equation}
    t_{\bullet} = (1 - 2) \times 0.6\times \frac{\eta \kappa c}{4\pi G} \simeq 23 - 46 \textrm{ Myr}.
\end{equation}
Note that this is comparable to the Salpeter e-folding timescale for a BH growing at its own Eddington rate, but in this case the BH mass can exponentiate multiple times.  In general we require mass inflow rates on the order of $\sim 0.1 - 1 M_{\odot}$ yr$^{-1}$ to form the quasi-star \citep{begelman06}, meaning that the BH will have grown to {\it few} $\times 10^{6-7} M_{\odot}$ in this time. 

Specifying the total mass of the quasi-star, the mass ratio $M_{\bullet}/M_{\star}$, and the luminosity enables the determination of other physical properties, including the radius $r_{\rm i}$ (which separates the saturated-convection and polytropic regions of the envelope), the quasi-star radius $R_{\star}$, and the temperature profile. This follows from the fact that $\xi_{\rm r}$ and $K_{\rm i}$ are functions only of the mass ratio, meaning that if we additionally specify $M_{\star}$, the mass interior to the saturated convection region, $M_{\rm i}$, can be determined. Equation \eqref{Kieq} then yields $r_{\rm i}$, which establishes the radius of the quasi-star, $R_{\star} = r_{\rm i}\xi_{\rm r}$. The temperature can be inferred by assuming that the total pressure consists of the sum of ideal gas pressure and radiation pressure:
\begin{equation}
    p = \frac{\rho k T}{\mu m_{\rm H}}+\frac{1}{3}aT^4,
\end{equation}
where $\mu$ is the mean molecular weight, for which we adopt $\mu = 0.6$ (valid for a completely ionized plasma that consists of 70\% hydrogen and 30\% helium by mass). 

\begin{figure*}
    \includegraphics[width=0.49\textwidth]{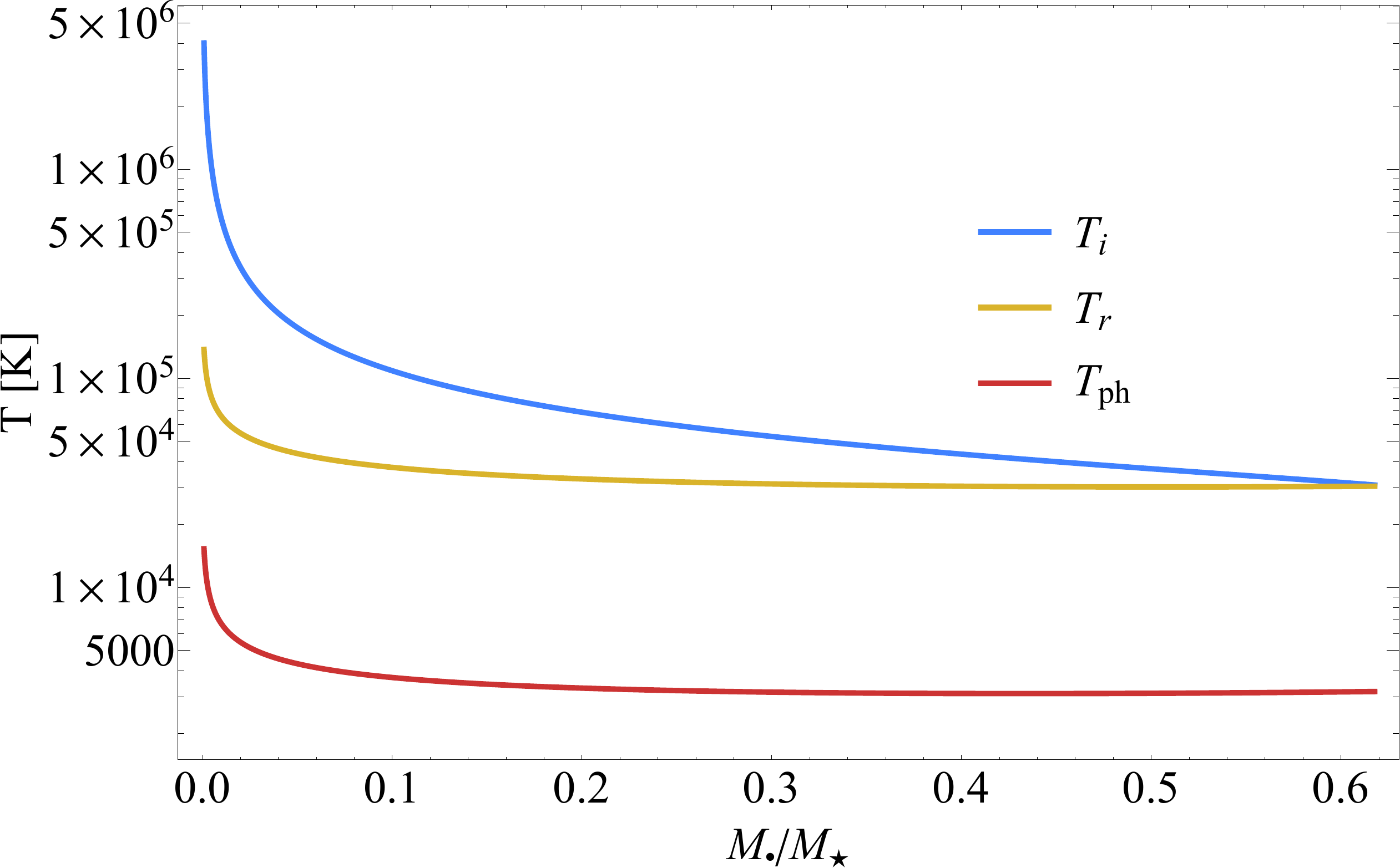}
  \includegraphics[width=0.48\textwidth]{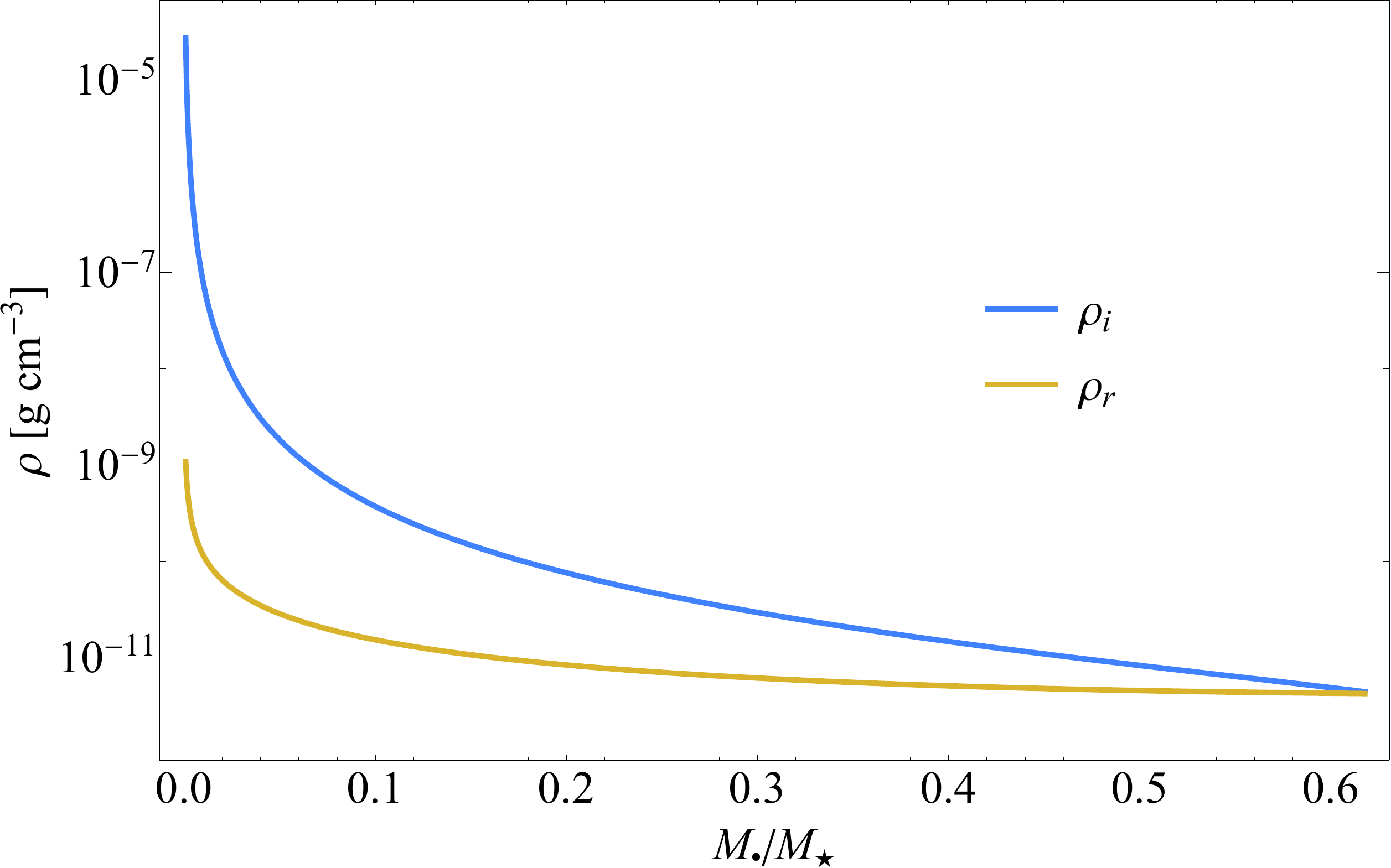} \hspace{-1in}
    \includegraphics[width=0.48\textwidth]{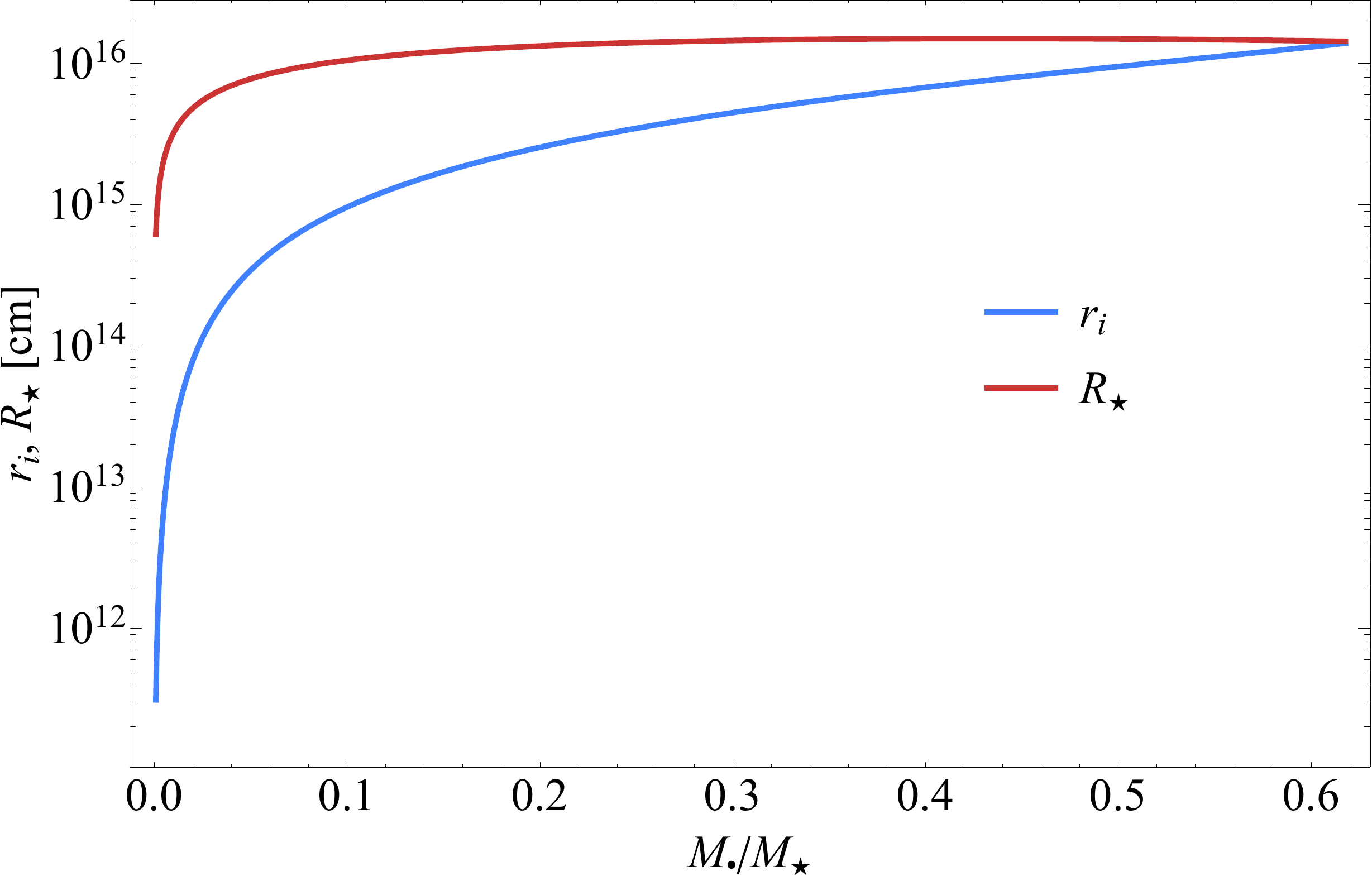}    
    \includegraphics[width=0.515\textwidth]{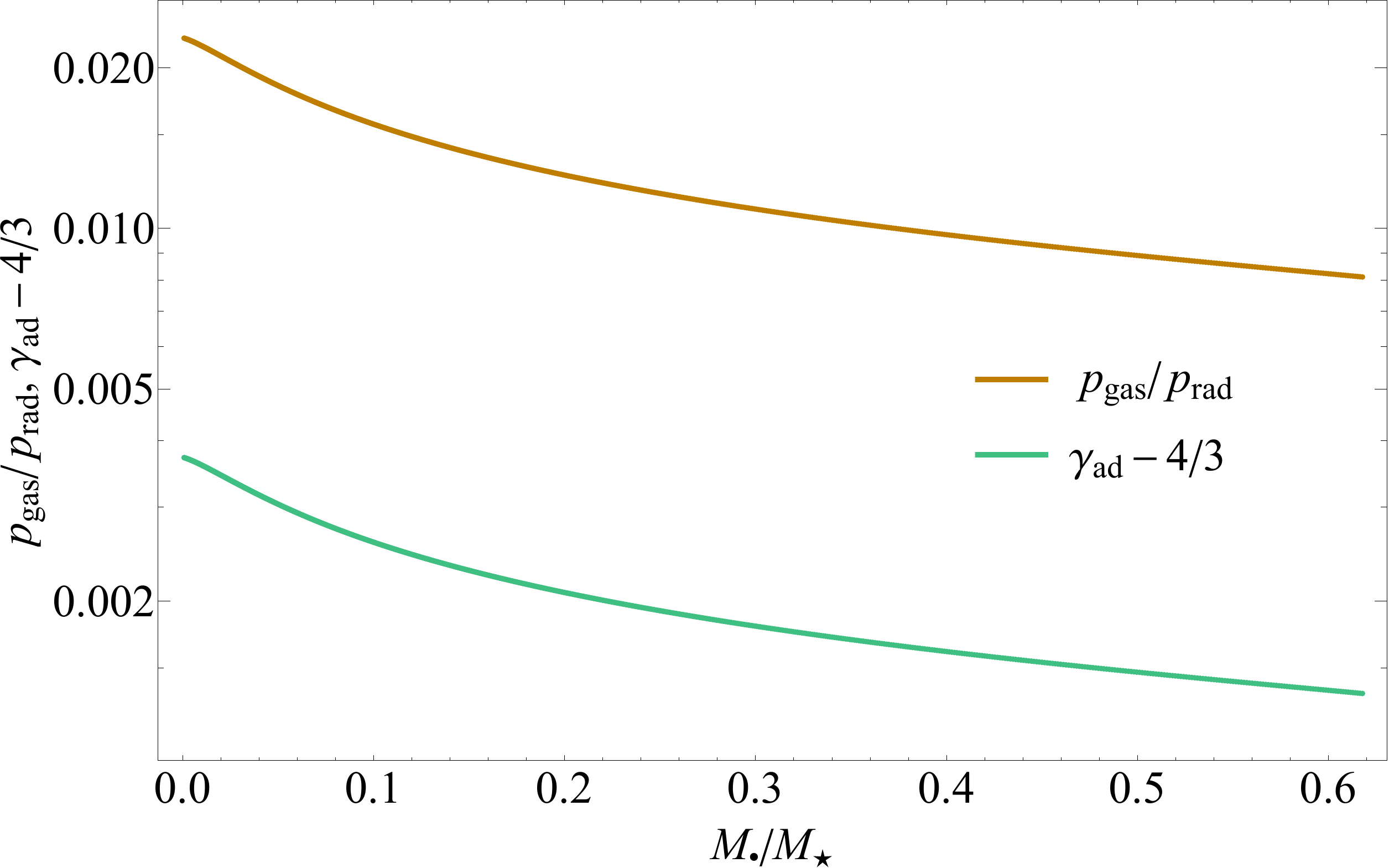}  
    \caption{Top-left: the temperature at the base of the polytropic region, $T_{\rm i}$, at the base of the radiative layer, $T_{\rm r}$, and the photospheric/effective temperature, $T_{\rm ph}$, as functions of the BH mass ratio. Top-right: the density at the base of the polytropic region, $\rho_{\rm i}$, and at the base of the radiative layer, $\rho_{\rm r}$. Bottom-left: The radius dividing the saturated-convection and polytropic regions, $r_{\rm i}$, and the radius of the quasi-star, $R_{\star}$. Bottom-right: the ratio of gas to radiation pressure and the adiabatic index relating the logarithmic pressure gradient to the logarithmic density gradient at constant entropy, i.e., $\gamma_{\rm ad} = \partial\ln p/\partial\ln \rho$. For all three panels the total quasi-star mass is fixed at $10^5 M_{\odot}$ and we let $\beta = 0.1$.} 
    \label{fig:T_of_ratio}
\end{figure*}

Figure \ref{fig:T_of_ratio} shows various physical quantities of the quasi-star for the fiducial case of $M_0 = 10^5M_{\odot}$, $\beta = 0.1$, $\dot{M} = 0$, as functions of $M_{\bullet}/M_{\star}$.  The top-left panel shows the temperature at $r_{\rm i}$, i.e., at the base of the polytropic envelope/edge of the saturated convection region, $T_{\rm i}$; at the base of the radiative zone, $T_{\rm r}$; and at the photosphere, $T_{\rm ph}$. The photospheric temperature is calculated as
\begin{equation}
    L = 4\pi\sigma R_{\star}^2T_{\rm ph}^4,
\end{equation}
i.e., assuming that the radiative layer is thin and the photospheric radius is $\sim R_{\star} = \xi_{\rm r}r_{\rm i}$. The top-right panel gives the densities at $r_{\rm i}$ and at the base of the radiative layer, and the bottom-left panel gives the radius of the saturated convection/polytropic boundary ($r_{\rm i}$) and the radius of the quasi-star ($R_{\star} = r_{\rm i}\xi_{\rm r}$). The bottom-right panel shows the ratio of gas to radiation pressure, $p_{\rm gas}/p_{\rm rad}$, and the difference between the adiabatic index, $\gamma_{\rm ad}$, that relates the logarithmic pressure gradient to the logarithmic density gradient, and $4/3$; if we let $p_{\rm gas}/p_{\rm rad} \equiv y$, the latter is
\begin{equation}
\begin{split}
    \gamma_{\rm ad} = \frac{\partial \ln p}{\partial \ln \rho}\bigg{|}_{\rm ad} &= \frac{\left(y+4\right)^2+3y\left(4+y/2\right)}{3\left(4+y/2\right)\left(y+1\right)} \\
    &= \frac{4}{3}+\frac{y}{6}+\mathcal{O}\left[y^2\right],
    \end{split}
\end{equation}
where the last equality results from a series expansion about $y = 0$.

From the top-left panel of this figure we conclude that, at early times and when the mass ratio is $\lesssim 0.1$, the temperature  is $\sim$ {\it few} $\times 10^{5-6}$ K at the base of the polytropic envelope, $\sim 10^{4-5}$ K at the transition to the radiative layer, and $\sim 5\times 10^3 - 10^4$ K at the photosphere. As the mass ratio grows, the temperatures all decline, reaching a $\sim$ mass-ratio-independent radiative and photospheric temperature of $T_{\rm r} \simeq 3\times 10^4$ K and $T_{\rm ph} \simeq 3\times 10^{3}$ K; high-mass quasi-star effective temperatures are thus comparable to those of red supergiants. Near the limiting mass when $M_{\star} \simeq M_{\rm i}$, the previous results imply that the photospheric temperature varies with $\beta$ and $M_{\star}$ as 
\begin{equation}
    T_{\rm ph} \simeq 3\times 10^{3}\left(\frac{\beta}{0.1}\right)^{-1/5}\left(\frac{M_{\star}}{10^5 M_{\odot}}\right)^{-1/20}\textrm{ K},
\end{equation}
i.e., the photospheric temperatures are very insensitive to both the choice of $\beta$ and the total quasi-star mass. The top-right panel indicates that the density at the base of the radiative layer also declines until $M_{\bullet}/M_{\star} \gtrsim 0.1$, above which it remains roughly constant at $\sim 10^{-12}$ g cm$^{-3}$, and the bottom-right panel shows that our assumption of a radiation-dominated gas is a very good one. These properties are all consistent with  the predictions of \citet{begelman08}.

The bottom-left panel of this figure shows that the saturated-convection radius, $r_{\rm i}$, depends strongly on the mass ratio, and increases rapidly as $M_{\bullet}/M_{\star}$ increases. This panel also shows that the radius of the quasi-star (equal to the radiative transition radius in our treatment) grows with increasing mass ratio when $M_{\bullet}/M_{\star} \lesssim 0.1$, but --- similarly to the temperature and density --- flattens to an approximately constant value (equal to $\sim 10^{16}$ cm) once the mass ratio satisfies $M_{\bullet}/M_{\star} \gtrsim 0.1$. We therefore see that BH growth is associated mainly with an increase in the size of the saturated-convection region: the BH eats the adiabatic quasi-star envelope from the inside out. The value of $\sim 10^{16} $ cm $\sim 10$ mpc is also slightly larger than the values obtained in the analytical model of \citet{begelman08}, who inferred radii a factor of $\sim 10$ smaller than this.  

In the preceding analysis, we have assumed that the radiative luminosity is produced by accretion onto the central BH. If the density profile of the saturated convection zone extends to near the gravitational radius of the BH, then we can check if this condition --- that all of the luminosity is produced via accretion --- can be achieved with a sub-luminal infall speed. Specifically, the accretion rate within the saturated convection region is $\dot{M} = 4\pi r^2\rho v$, where $v$ is the net infall velocity that we are assuming is small (relative to the escape speed). Using our scalings for various quantities above and assuming $\xi \ll 1$, the accretion rate becomes
\begin{equation}
    \dot{M} = 4\pi r^2\rho v = \frac{9}{4\beta\sqrt{2}}\frac{L}{c^2}N^{3/2}\frac{v}{c}.
\end{equation}
Here we set the physical (i.e., not scaled by $r_{\rm i}$) radius at which we are evaluating $\dot{M}$ to $r = N\times GM_{\bullet}/c^2$, and used the asymptotic density profile at $\xi \ll 1$ as derived in Section \ref{sec:inner}. When $N \sim 1$ we require the velocity to become an order-unity fraction of the speed of light if accretion provides the luminosity. Setting $L = \eta\dot{M}c^2$ and rearranging then shows that the velocity is
\begin{equation}
    \frac{v}{c} = \frac{4\sqrt{2}}{9}\frac{\beta}{\eta}N^{-3/2}.
\end{equation}
If $\beta \simeq \eta \simeq 0.1$, then $N \simeq$ {\it few} shows that this is a self-consistent assumption --- the infall speed at radii comparable to the gravitational radius must be $v/c \simeq$ {\it few} $\times 0.1$ to enable BH accretion to power the envelope when the saturated-convection region extends to radii comparable to the gravitational radius of the BH.

\section{Summary and Conclusions}
\label{sec:summary}
In this paper we constructed a model that applies to the intermediate evolutionary stage of quasi-stars, which are supercritically accreting BHs surrounded by massive ($M_{\star} \gtrsim 10^5M_{\odot}$) gaseous envelopes, with the envelope energetically sustained by the BH accretion. As described in \citet{begelman06, begelman08}, this stage of BH growth should be present alongside any of the ``direct collapse'' models of early galaxy formation, which are consistent with (and may be required to produce) the recent JWST results that show evidence for high-mass galaxies and quasars at redshifts $z \gtrsim 10$ (e.g., \citealt{bogdan24}). In Section \ref{sec:quasi-stars-low} we argued that the boundary condition used by \citet{begelman08} and \citet{ball11, ball12} --- which effectively serves to determine the Bondi radius of the BH in terms of the quasi-star properties --- and the corresponding BH mass limit of $M_{\bullet}/M_{\star} \sim 0.017$ is not physical. Instead, we suggested that once the BH mass grows to $\sim 1\%$ of the mass of the star, it is significant enough to modify the internal structure of the quasi-star envelope, pushing the hydrostatic region to smaller radii and invalidating the use of the Bondi condition. Rather than implying a limit on BH growth, this bound simply suggests that the energy source responsible for sustaining the overlying envelope is concentrated at smaller radii, closer to the BH.

Based on this reasoning, in Section \ref{sec:quasi-stars-moderate} we proposed that the large energy generation rate near the BH likely drives the convective flux in the inner regions to its theoretical maximum, $F_{\rm con, max} \simeq p c_{\rm s}$. Demanding that the convective flux transport all of the energy then yields an effective equation of state between the pressure and density (Equation \ref{peos}), and in the limit that the flow is non-self-gravitating gives the radially dependent density and pressure profiles $\rho \propto r^{-1/2}$ and $p \propto r^{-3/2}$ --- identical to the spherically averaged profiles of the convection-dominated accretion flows described in \citet{quataert00}. By matching the entropy, pressure, and mass between the saturated-convection inner region and $\sim$ adiabatic/polytropic outer envelope (within which a small entropy gradient presumably transports the energy), we showed that the total solutions were functions only of the mass ratio $M_{\bullet}/M_{\star}$ (see Figures \ref{fig:mass_of_r} and \ref{fig:mbullet_saturated}). These solutions permit a maximum black hole mass of $M_{\bullet}/M_{\star} \simeq 62$\%, with almost all of the remaining $\sim 38$\% contained in the saturated convection region, allowing the BH mass to reach order-unity fractions of the mass of the star. 

In Section \ref{sec:radii} we analyzed the physical properties (e.g., radius and effective temperature) of the quasi-star, which --- because the solutions in Section \ref{sec:quasi-stars-moderate} are completely self-similar and depend only on $M_{\bullet}/M_{\star}$ --- require specification of the total mass of the star, the BH mass ratio, and the luminosity. We constrained the latter by assuming that the quasi-star radiates at the Eddington limit corresponding to the total mass, which is a good approximation if the star is in hydrostatic equilibrium in the radiative surface layers and radiation-pressure dominated. The physical properties that result are similar to those of previous works (e.g., \citealt{begelman08}), and if the quasi-star accretes from its surroundings at a rate of $0.1 - 1 M_{\odot}$ yr$^{-1}$, the BH can grow to $\sim 10^{6-7}M_{\odot}$ in $\sim 20-40$ Myr. The quasi-star picture, and specifically the model we developed here, is therefore capable of producing SMBHs in the early Universe, consistent with recent JWST observations.

Our analysis was simplified from the standpoint that we did not consider the structure of the radiative and outermost layers of the star. Instead, we assumed that the radiative region was geometrically thin, with its base (i.e., the radius at which radiative diffusion takes over as the dominant mode of energy transport) determined by where convection is no longer capable of transporting the total luminosity. Accounting for opacity effects and calculating the photospheric radius self-consistently could lead to regions of super-Eddington luminosity within the radiative layer, as was found in \citet{begelman08}. It may therefore be the case that opacity effects limit the lifetime of the quasi-star envelope --- and thus the maximum achievable BH mass --- to times before order-unity mass ratios are reached, should super-Eddington feedback impart significant momentum to the outer envelope and drive a wind; the effects of a radiatively driven wind were considered by \citet{fiacconi16} (\citealt{fiacconi17} also analyzed the impact of rotation). It is also not clear how the interaction between this outflow and the infalling gas, which is ultimately responsible for establishing the initial mass and subsequent growth of the quasi-star, would modify this simple picture. 

The accretion rates onto the central BH are supercritical by orders of magnitude, which validates the assumption of our model that convection transports the accretion energy outward, but these conditions are also conducive to the formation of outflows/jets from near the BH (the possibility of jets from quasi-stars was described in \citealt{czerny12}). If a jet is formed and provided that there is not substantial feedback between the jet and the overlying envelope, this could provide another route by which the energy can escape the system, which may be necessary if realistic opacities and convective efficiencies (i.e., the value of $\beta$ that we canonically set to 0.1; see also \citealt{begelman08}) drive the photospheric luminosity to super-Eddington values. Since the density and pressure profiles decline as $\rho \propto r^{-1/2}$ and $p \propto r^{-3/2}$ in the deep interior, the jet can be readily collimated by the ambient medium through oblique shocks \citep{levinson00, bromberg07, zakamska08, kohler12} or local and viscous effects that arise from the intense radiation field itself \citep{coughlin20}, or by the surrounding cocoon of shocked material produced as a byproduct of the advancement of the jet through the star \citep{lazzati05, bromberg11, levinson13}.

On the other hand, it is not clear if the jet will successfully propagate to the surface of the quasi-star prior to being choked or pinched off from the surrounding and high-pressure cocoon that is created as a byproduct of shocked ambient and shocked jet material \citep{begelman89}. For example, \citet{matzner03} (see also \citealt{marti94}) analytically analyzed the propagation of a causally connected jet through the interior of a star, concluding that the jet head (or at least the forward shock and contact discontinuity) could only be relativistic for compact progenitors, i.e., Wolf-Rayets and blue supergiants; see also the numerical simulations by \citet{morsony07, lazzati09, nagakura11}. In highly extended quasi-star envelopes, it may therefore be that the jet does not remain relativistic and does not provide an efficient exhaust route, but is instead responsible for depositing the energy at larger radii within the envelope and ultimately leading to its destruction. Furthermore, if the jet is responsible for carrying away the majority of the accretion energy, it may not be necessary to enforce the saturated convection equation of state employed here. Instead, the inner regions --- and even the flow on larger scales --- may more closely resemble weakly bound and quasi-spherical, zero-Bernoulli accretion flows (ZEBRAs) described in \citep{coughlin14}, although the structure would be modified by the self-gravitating nature of the gas. A more detailed investigation is necessary to understand and quantify the nature of jet propagation in quasi-stars and the corresponding impact on the inner disc and envelope structure.

{\citet{begelman08} and \citet{ball11} numerically determined quasi-star structures in the framework of stellar evolution, within which the system evolves through a sequence of steady states, and this assumption is implicit in our discussion of the physical properties in Section \ref{sec:radii}. Unlike the case where the computational domain is truncated at the Bondi radius, at which point boundary conditions were imposed by both of these previous studies, the inner region for the model described here --- or at least the pressure, temperature, and mass at the inner radius --- would have to be determined self-consistently alongside the envelope structure, as the properties of the inner region depend on the bulk properties of the quasi-star. It is only because we assumed a polytropic and radiation-pressure-dominated equation of state and terminated the solutions at the radius where convection becomes inefficient relative to radiative diffusion (see Section \ref{sec:quasi-stars-moderate}) that we were able to construct self-similar solutions that depend only on a single, dimensionless parameter ($K_{\rm i}$ in our notation). Relaxing these assumptions and allowing for, e.g., finite gas pressure, radiative diffusion (and corresponding opacity effects), and non-adiabatic convection would quantify the impact of the radiative layer and yield more accurate quasi-star structures, which could differ non-trivially from the analytic solutions explored here. On the other hand, neglecting hydrodynamic effects and the presence of a finite radial velocity (let alone multi-dimensional effects) remains questionable, owing to the fact that radiatively driven winds may be generated (as described in the preceding paragraphs and by \citealt{fiacconi16}), the structures themselves are likely only marginally stable (see the ensuing paragraph), and the quasi-star is accreting from its surroundings. In one dimension and for small mass ratios, including these effects is likely to be challenging, owing to the large disparity in spatial scales between the inner region and the quasi-star radius (see Figure \ref{fig:T_of_ratio}), though it could become more feasible as the mass ratio grows and the computational domain is less expansive.}

Finally, the question of the stability of quasi-stars is an important one, as radiation-dominated stars with polytropic and adiabatic indices $\simeq 4/3$ are prone to dynamical (i.e., Chandrasekhar-) instability. Additionally, very massive stars may be prone to a thermal instability that is distinct from the Chandrasekhar instability and more akin to the Jeans instability \citep{thompson08, thompson10}. While a stability analysis of the solutions described here is outside the scope of the present investigation, we argue that the existence of the saturated-convection region likely has a stabilizing effect at small mass ratios, owing to the fact that $p \propto r^{-1/2} \propto \rho^{3}$. The relationship $p \propto \rho^3$ is a $\gamma$-law equation of state with a stiff polytropic exponent, implying that the pressure in the inner regions of the quasi-star will change more rapidly than the self-gravitational field in the presence of a perturbation compared to a pure $4/3$-polytrope, thus stabilizing the star (see \citealt{kundu21} for a more rigorous demonstration of the validity of this argument). We leave a more in-depth analysis of quasi-star stability, and specifically the solutions described here, to future work.

\acknowledgements
{We thank the anonymous referee for useful suggestions that improved the manuscript.} E.R.C.~acknowledges support from the National Science Foundation through grant AST-2006684 and from NASA through the Astrophysics Theory Program, grant 80NSSC24K0897.  M.C.B.~acknowledges support from  NASA Astrophysics Theory Program grants NNX17AK55G and 80NSSC22K0826, and NSF Grant AST-1903335.

\bibliographystyle{aasjournal}


\end{document}